\documentclass[twocolumn,superscriptaddress,preprintnumbers,amsmath,amssymb,prl]{revtex4-1}

\newcommand{\IMSS}{Muon Science Laboratory, Institute of Materials Structure Science, High Energy Accelerator Research Organization (KEK), Tsukuba, Ibaraki 305-0801, Japan}
\newcommand{\IMSSN}{Neutron Science Division, Institute of Materials Structure Science, High Energy Accelerator Research Organization (KEK), Tsukuba, Ibaraki 305-0801, Japan}
\newcommand{\Sokendai}{Department of Materials Structure Science, The Graduate University for Advanced Studies (Sokendai), Tsukuba, Ibaraki 305-0801, Japan}
\newcommand{\MCES}{Materials Research Center for Element Strategy (MCES), Tokyo Institute of Technology, Yokohama, Kanagawa 226-8503, Japan}
\newcommand{\KyotoU}{Department of Energy and Hydrocarbon Chemistry, Graduate School of Engineering, Kyoto University, Nishikyo-ku, Kyoto 615-8510, Japan}

\usepackage[dvipdfmx]{graphicx}
\usepackage{dcolumn}
\usepackage{multirow}
\usepackage{bm}
\usepackage[hypertex,colorlinks=true,linkcolor=black,citecolor=blue,filecolor=blue,urlcolor=blue,setpagesize=false,nesting=true]{hyperref}

\usepackage[utf8]{inputenc}
\usepackage[T1]{fontenc}
\usepackage{mathptmx}

\begin{document}
\title{Local spin dynamics in geometrically frustrated Mo pyrochlore antiferromagnet Lu$_2$Mo$_2$O$_{5-y}$N$_2$ 
}
\author{S. K. Dey}
\affiliation{\IMSS}
\author{K. Ishida}
\affiliation{\KyotoU}
\author{H. Okabe}
\affiliation{\IMSS}
\author{M. Hiraishi}
\affiliation{\IMSS}
\author{A. Koda}
\affiliation{\IMSS}\affiliation{\Sokendai}
\author{T. Honda}\affiliation{\Sokendai}
\affiliation{\IMSSN}
\author{J. Yamaura}
\affiliation{\MCES}
\author{H. Kageyama}
\affiliation{\KyotoU}
\author{R. Kadono}\email{ryosuke.kadono@kek.jp}
\affiliation{\IMSS}\affiliation{\Sokendai}

\begin{abstract}
The magnetic ground state of oxynitride pyrochlore Lu$_2$Mo$_2$O$_{5-y}$N$_2$, a candidate compound for the quantum spin liquid ($S=1/2$, Mo$^{5+}$), was studied by muon spin rotation/relaxation experiment.   In contrast to Lu$_2$Mo$_2$O$_7$ ($S=1$, Mo$^{4+}$) which exhibits the spin-glass behavior with a freezing temperature $T_g\simeq16$ K, no such spin freezing or long range magnetic order was observed down to 0.3 K. Moreover, two separate magnetic domains were detected below $\sim$13 K, which were characterized by differences in spin dynamics. The first is the ``sporadic'' spin fluctuation seen in frustrated antiferromagnets, where the amplitude of the hyperfine fields suggests that the excitation comprises a local cluster of unpaired spins. The other is rapid paramagnetic fluctuation, which is only weakly suppressed at low temperatures. In place of the paramagnetic fluctuation, the volume fraction for the sporadic fluctuation steadily increases with decreasing temperature, indicating the formation of an excitation gap with a broad distribution of the gap energy involving null gap.
\end{abstract}

\maketitle
\section{Introduction}
Geometrically frustrated magnets have been quite extensively studied in the quest of exotic magnetic ground states \cite{Diep:05,Lacroix:11}. The local spins in these magnetic lattices are arranged on the vertices of the corner shared triangular or tetrahedral units so that the pairwise satisfaction of minimizing energy is not achieved under isotropic antiferromagnetic interaction with the nearest neighbor (nn), leading to the macroscopic degeneracy of ground states. Moreover, quantum fluctuations associated with small spin quantum numbers can produce quantum spin liquid (QSL) states in which there is no magnetic long-range order or spontaneous symmetry breaking down to zero-temperature limit ($T\rightarrow0$) \cite{Lee:08,Balents:10,Zhou:17}. 

Understanding the nature of QSLs requires detailed knowledge of the low-lying elementary excitations for $T\rightarrow0$, particularly the presence or absence of an excitation gap. This information has immediate implications for the spin correlations of the ground state as well as the correlation length scale of the QSL.
For example, in one-dimensional (1D) spin-1/2 Heisenberg chains, the elementary excitations are gapless spinons (charge-less spin-1/2 quasiparticles) with a linear energy dispersion and a power-law decay of the spin correlation \cite{Faddeev:81}, whereas in the integer spin case such excitations are gapped \cite{Haldane:83}. 
While there have been several two-dimensional (2D) materials that exhibit gapless or nearly gapless QSL-like properties, candidates in three-dimensional (3D) materials are relatively scarce \cite{Gardner:99,Fennell:12,Clark_PRL,3D_QSL_Smith_PRB, 3D_QSL_Gao2019, 3D_QSL_Chillal2020, Khuntia_PRL}. Among these, the model system of quantum spin pyrochlore Heisenberg antiferromagnets with geometric frustration has been regarded a promising candidate for 3D QSL states \cite{3D_QSL_PRL_Canals_theory, 3D_QSL_pyrochlore_lattice_theory, 3D_QSL_pyrochlore_Normad}.

Lu$_2$Mo$_2$O$_5$N$_2$ is an insulating oxynitride compound derived from molybdate pyrochlore antiferromagnet Lu$_2$Mo$_2$O$_7$. The substitution of oxygen (O$^{2-}$) with nitrogen  (N$^{3-}$) changes the valence of Mo from Mo$^{4+}$ (4$d^2$) to Mo$^{5+}$ (4$d^1$), which is presumed to constitute spin-1/2 Heisenberg antiferromagnet at nominal composition.   The bulk magnetic susceptibility exhibits no sign of spin freezing or long-range magnetic order down to 2 K \cite{Clark_PRL}.  The magnetic specific heat shows a linear temperature dependence for 0.5--30 K, suggesting a spinon Fermi surface \cite{Irkhin:90}.   Inelastic neutron scattering (INS) experiment has revealed a broad dynamic structure factor at 1.5 K due to strong spin fluctuations \cite{Clark_PRL}. These results imply that the QSL state with a negligibly small gap is realized in Lu$_2$Mo$_2$O$_5$N$_2$.  Density functional theory (DFT) calculations predicted that the compound is well described by the Heisenberg model with exchange parameters up to third nearest neighbors and is intrinsically paramagnetic down to $|\theta_{\rm CW}|$/100 by geometrical frustration (where $\theta_{\rm CW} = -121$ K is the Curie-Weiss temperature)  \cite{Iqbal}. 

Meanwhile, the effect of bond randomness (disorder) on the QSL states has been studied theoretically in a variety of 2D systems, where it was predicted that gapless nonmagnetic random singlet states appear when randomness exceeds a critical value \cite{Watanabe:14,Kawamura:14,Kimchi_PRX,Lu_Liu_PRX,H_Wu_PRB,Kawamura:19}.  This prediction is not self-evident, as it is believed that in pure systems, the slightest disorder breaks the QSL state and induces a spin freezing, as seen in the example of Y$_2$Mo$_2$O$_7$ \cite{Gingras:97,Booth:00,Keren:01,Saunders:07}.
The disorder-induced QSL state can be visualized as a random arrangement of spin-singlet dimers between spins with varying distances. Unpaired spins that fail to form a singlet dimer migrate in the sea of singlet dimers by spinon-like orphan-spin excitations, imparting a linear temperature dependence to the low-temperature specific heat. In a disorder-free system, spinon excitations are expected to propagate ballistically, while orphan-spin motions are expected to be essentially diffusive due to the bond randomness.

One of the first 3D systems to be scrutinized for the effects of bond rondamness was Lu$_2$Mo$_2$O$_5$N$_2$, where a situation qualitatively similar to that of 2D systems is predicted \cite{Uematsu_PRL}. The gapless behavior with a Curie law-like tail in the temperature range of 0.5--2 K is consistent with the existence of orphan spins in the spin singlet state in Lu$_2$Mo$_2$O$_5$N$_2$.  
However, microscopic information on the spin dynamics in Lu$_2$Mo$_2$O$_5$N$_2$ is so far limited to the INS result at 1.5 K \cite{Clark_PRL}.

Muon spin rotation and relaxation ($\mu$SR) is a powerful method to study local spin dynamics in a unique time window of $10^{-5}$--$10^{-9}$ s, complementary to other microscopic spin probes such as nuclear magnetic resonance and INS.  In particular, it has been used to reveal peculiar spin fluctuations in geometrically frustrated magnetic materials \cite{Lacroix:11}.  While bulk properties such as specific heat are sensitive to all kinds of excitations, implanted muons selectively probe magnetic excitations via the local spin fluctuation. Here, we report on the $\mu$SR experiment of Lu$_2$Mo$_2$O$_{5-y}$N$_2$ over the temperature range down to 0.3 K. It reveals that two distinct magnetic domains emerge with decreasing temperature. One is a domain characterized by the ``sporadic'' spin dynamics as reported earlier for 
frustrated antiferromagnets \cite{Cho,Khuntia_PRL,Undecouplable_Gaussian_FUKAYA, Undecouplable_Gaussian_Aczel_PRB,Undecouplable_Gaussian_PRB_Yoon,Uemura_SCGO,Bono,Undecouplable_Gaussian_Lee_PRB,Uemura_SCGO,Bono,Bono_2005,Bono_2006}. The amplitude of the hyperfine fields $\delta$ implies that this excitation consists of a local cluster of unpaired  ($S=1/2$) spins, and there is no significant decrease in $\delta$ as expected for isolated spinon excitations.   The other exhibits fast paramagnetic fluctuations that are only weakly suppressed at low temperatures.

The presence of an oxygen deficit ($y\simeq0.4$) in our sample suggests that these two domains correspond to two different Mo valence states, with the sporadic spin fluctuations hosted by the Mo$^{5+}$-rich domains, and the other in the Mo$^{4+}$-rich domains.  The increase in the volume fraction $x$ for the sporadic fluctuation with decreasing temperature suggests that this ground state is accompanied by a finite excitation gap. However, the temperature dependence of $x$ implies the existence of a broad distribution of gap energies, including a null excitation gap, which can be interpreted as an effect due to bond randomness.  Furthermore, the predominance of paramagnetic fluctuation in the secondary domains suggests that the bond randomness serves to suppress the spin freezing in the Mo$^{4+}$ domains.

\section{Experimental Details}
Polycrystalline oxynitride samples were synthesized by solid phase reactions and ammonolysis techniques as described elsewhere \cite{Clark_PRL}. To investigate Mo valence, their quality was evaluated by dc magnetic susceptibility, CHN elemental analysis, synchrotron X-ray diffraction, powder neutron diffraction (PND), and Mo $K$-edge X-ray absorption near edge structure (XANES); details are found in Appendix A.
These analyses suggest that the presence of O deficiency reduced the average valence to Mo$^{4.57(8)+}$, corresponding to $y=0.43(8)$. Therefore, bond randomness in this sample is presumed to be induced by both O deficiency and N substitution. 

 The $\mu$SR measurements were performed using the S1 instrument ARTEMIS in MLF, J-PARC, where a nearly 100\% spin-polarized pulsed muon beam ($\mu^+$, with the full width at half-maximum of 100 ns and a momentum of 27 MeV/c) was transported to the sample.  A detailed experimental account is given in Appendix B. For controlling sample temperature, $^4$He flow-type and $^3$He-type cryostats were employed over the region 5--298 K and 0.3--30 K, respectively.

\section{Results}

\subsection{Muon Site}
\begin{figure}[t]
	\begin{center}
		\includegraphics[width=0.45\textwidth]{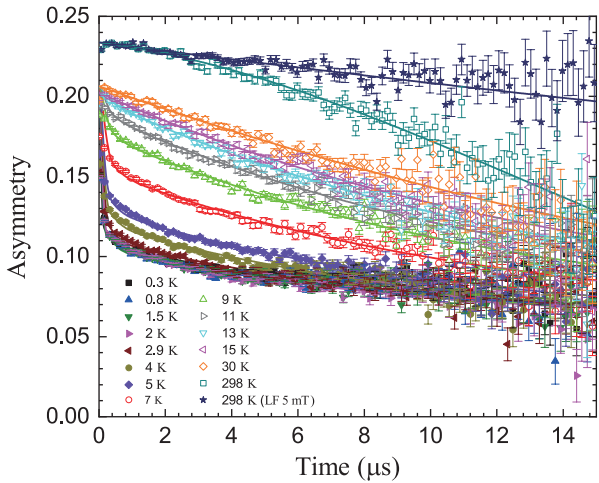}
	\end{center}
	\caption{Temperature dependence of the zero field (ZF) $\mu$SR time spectra in Lu$_2$Mo$_2$O$_{5-y}$N$_2$. The spectra at 298 K under ZF and a longitudinal field of 5 mT were measured using $^4$He flow-type cryostat, and thereby the initial asymmetry is slightly different.  Solid curves represent the best fits obtained by the least-square analysis (see text).}
	\label{ZF_spectra_Heliox}
\end{figure}

We quantitatively evaluated the spin fluctuation in $\mu$SR experiment by first measuring the detailed change with temperature under zero field (ZF) conditions and then measuring the detailed longitudinal field (LF) dependence at characteristic temperatures where a distinct change is observed in the ZF spectrum.
Fig.~\ref{ZF_spectra_Heliox} shows examples of ZF-$\mu$SR time spectra [the time-dependent $\mu^+$-$e^+$ decay asymmetry, $A(t)$] in the temperature range of 0.3--30 K. 
There is little change at temperatures above 15 K, and the depolarization is mainly due to the hyperfine fields ($H_{\rm n}$) exerted from nuclear magnetic moments.  The lineshape is well described by the product of the Gaussian Kubo-Toyabe (GKT) relaxation function, $G_{z}^{\rm KT}(t; H_L,\delta,\nu)$, and a slow exponential damping due to the fluctuation of paramagnetic Mo moments, 
  \begin{equation}
 	A(t) = A_0G_{z}^{\rm KT}(t; H_L,\delta_{\rm n},\nu_{\rm n})\cdot \exp(-\lambda^\prime t) +A_{\rm b},
 	\label{eqn:G_KT}
 \end{equation}
 where $H_L$ is the applied longitudinal field in the direction of the initial muon polarization, $\delta_{\rm n}$ is the linewidth determined by the distribution of $H_{\rm n}$, and $\nu_{\rm n}$ is the fluctuation rate of $\delta_{\rm n}$ \cite{Hayano_PRB}.  
In the case of zero external field and $\nu_{\rm n}\ll\delta_{\rm n}$ (i.e., static $H_{\rm n}$), we have $G_{z}^{\rm KT}(t; 0,\delta_{\rm n},\nu_{\rm n})\simeq\exp(-\delta_{\rm n}^2t^2)$.
 (For more details on the hyperfine interactions between the muon and nuclei  and resulting GKT relaxation functions, see Appendix C.)

In the paramagnetic phase, the depolarization due to relatively fast fluctuations of the hyperfine fields from electron magnetic moments is given by $\exp(-\lambda^\prime t)$, where $\lambda^\prime$ is determined by the hyperfine fields from unpaired Mo spins.   As discussed below, the GKT function is also employed to describe the muon depolarization due to the contribution of Mo electron moments, where $\delta_{\rm n}$ and $\nu_{\rm n}$ are replaced by those of unpaired Mo electrons, $\delta$ and $\nu$. [The exponential decay in Eq.~(\ref{eqn:G_KT}) corresponds to the approximation of the GKT function for the condition $\nu\gg\delta$; see Eqs.~(\ref{Gexp}), (\ref{BPP}).]

As seen in Fig.~\ref{ZF_spectra_Heliox}, the lineshape of the ZF spectrum at 298 K is Gaussian-like ($\lambda^\prime<\delta_{\rm n}$), and the magnitude of $\delta_{\rm n}$ is reliably estimated as $\delta_{\rm n}=0.0458(9)$ MHz by curve fitting combined with the LF spectrum. By comparing this value with the calculated $\delta_{\rm n}$, the muon site can be determined.  

\begin{figure}[t]
	\begin{center}
		\includegraphics[width=0.47\textwidth]{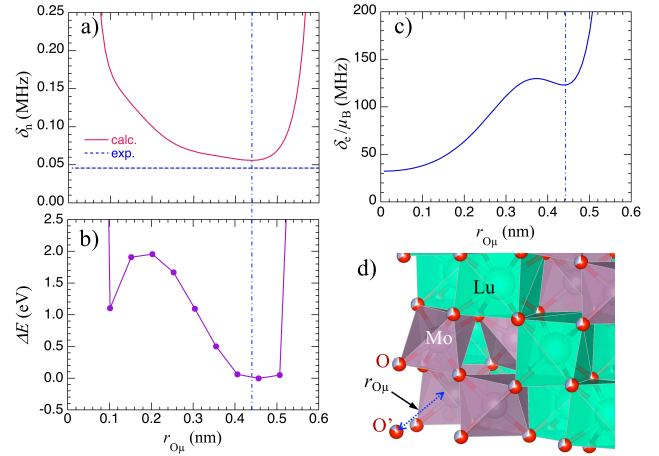}
	\end{center}
	\caption{a) The linewidth for the nuclear random local fields,  b) total energy for the interstitial hydrogen plus Lu$_2$Mo$_2$O$_{5}$N$_2$ system (with the origin at the minimum), and c) the hyperfine parameter due to the magnetic dipole moment of Mo electrons, which are calculated at a distance $r_{\rm O\mu}$ from an oxygen atom as defined in d). The crystal structure is depicted using VESTA~\cite{Vesta}.}
	\label{Musite}
\end{figure}

    Figure \ref{Musite} shows a summary of the physical parameters calculated at a distance $r_{\rm O\mu}$ from an oxygen atom (O') in the [111] direction, with the minimum total energy included on this axis.  As shown in Fig.~\ref{Musite}(a), the calculated $\delta_{\rm n}$ using Eq.~\ref{delta_n} (in Appnd.~C) is closest to the experimental value at $r_{\rm O\mu}=0.44$ nm, suggesting that this location is the only candidate site. It is often the case that muons form OH bonds with ligand O and are located at a site $r_{\rm O\mu}\simeq0.1$ nm from oxygen.  However, it is obvious in Fig.~\ref{Musite}(a) that the calculated $\delta_{\rm n}$ at this position is more than three times the experimental value, 
which experimentally rules out the possibility of the O$\mu$ bonding site. Meanwhile, since the calculated $\delta_{\rm n}$ showed a broad tail towards oxygen site, we used the DFT calculation to refine the muon location. As a result, we found that the candidate site is consistent with that of the minimal total energy shown in Fig.~\ref{Musite}(b) (see Appnd.~B for the details). From these we conclude that the muon occupies a site near the center of the tetrahedron consisting of four nn Mo ions at the corners. The hyperfine parameter $\delta_{\rm e}$ (per Bohr magneton) calculated for the Mo electron spins  using Eq.~\ref{delta_e} (in Appnd.~C) is shown in Fig.~\ref{Musite}(c), from which we obtain $\delta_{\rm e}/\mu_B = 122.9$ MHz/$\mu_B$ at the muon site, which is used to evaluate the effective Mo moment size at lower temperatures.

\subsection{Two types of magnetic domains inferred from ZF/LF-$\mu$SR spectra}

  Below $\sim$15 K, fast depolarizing component appears in the early time region ($t\leq1$ $\mu$s), implying that the slowing down of the spin fluctuations sets in. However, the absence of spontaneous oscillatory signals in these time spectra indicates that the long-ranged magnetic order does not occur down to 0.3 K.  In this component, the depolarization rate increases with decreasing temperature down to 2 K, below which there is no change in the depolarization rate.  Whether this is due to spin freezing can be determined from the LF-$\mu$SR spectrum.

\begin{figure}[t]
  	\begin{center}
  		\includegraphics[width=0.45\textwidth]{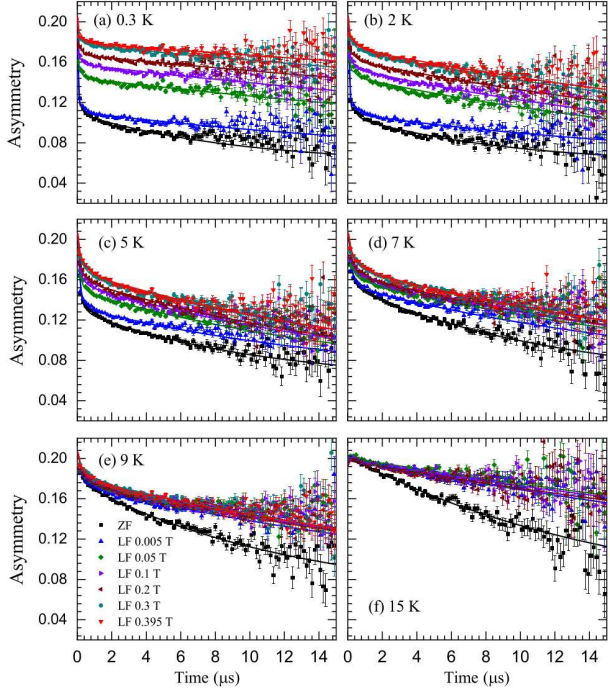}
  	\end{center}
  	\caption{Examples of $\mu$SR time spectra observed at typical temperatures 
  			in Lu$_2$Mo$_2$O$_{5-y}$N$_2$ under various longitudinal fields: a) 0.3 K, b) 2 K, c) 5 K, 
			d) 7 K, e) 9 K, and f) 15 K. Solid curves are the best fits by the least-square analysis using the model described in the text.}
  	\label{LFtspec}
  \end{figure}

Figure \ref{LFtspec}  shows typical examples of the LF-$\mu$SR time spectra over a temperature range from 0.3 to 15 K, obtained above 5 mT to quench the contribution of nuclear magnetic moments [i.e., $G_{z}^{\rm KT}(t; H_L,\delta_{\rm n},\nu_{\rm n})\simeq1$ for  $\gamma_\mu H_L\gg\delta_{\rm n}$ in Eq.~(\ref{eqn:G_KT}), $\gamma_\mu=2\pi \times 135.53$ MHz/T is the gyromagnetic ratio of muon spin] and extract the electronic contribution [$\exp(-\lambda^\prime t)$]. One of the major challenges in analyzing these time spectra was to find a realistic physical model to describe the temperature and magnetic field dependence of $G_z(t)$ coherently. From Fig.~\ref{LFtspec}, it was obvious that the single exponential decay would not satisfy this objective.  It was also found that introducing a conventional GKT function (with $\delta_{\rm n}$ replaced by that for the electronic hyperfine fields $\delta$) did not improve the situation. (It should be emphasized that there is a clear rationale for expecting a description of $G_{z}(t)$ by the GKT function here, as indicated in Appnd.~C.) In the following, we show that all these spectra can be reproduced coherently by considering two channels of depolarization originating from two different spin dynamics. One is due to the ``sporadic" field fluctuation described by the  ``undecouplable'' Gaussian KT (uGKT) function \cite{Uemura_SCGO} and the other is due to the fluctuation modeled by the conventional Markovian process.

The uGKT function is known to describe the muon depolarization due to sporadic field fluctuations seen in various spin-1/2 frustrated magnets including CuAl$_2$O$_4$ \cite{Cho}, PbCuTe$_2$O$_6$ \cite{Khuntia_PRL}, SrCu$_2$(BO$_3$)$_2$ \cite{Undecouplable_Gaussian_FUKAYA, Undecouplable_Gaussian_Aczel_PRB}, and PbCu$_3$TeO$_7$ \cite{Undecouplable_Gaussian_PRB_Yoon}.  Interestingly, it has been reported that some systems with $S\ge1$ such as Cr-kagome bilayer compounds, SrCr$_{9p}$Ga$_{12-9p}$O$_{19}$ [SCGO(p)] and Ba$_2$Sn$_2$ZnGa$_{10-7p}$Cr$_{7p}$O$_{22}$  [BSZCGO(p)] \cite{Uemura_SCGO,Bono}, and  YBaCo$_4$O$_7$ \cite{Undecouplable_Gaussian_Lee_PRB} also exhibit spin fluctuations that are well described by the uGKT function.  The function is derived by introducing a scaling factor $f$ ($<1$) in the time domain to the conventional dynamical GKT function as $G_z^{\rm KT}(ft; H_L,\delta, \nu)$ \cite{Uemura_SCGO}. The physical interpretation is that the muons feel the hyperfine field $\delta$ for $ft$ due to the sporadic appearance of unpaired spins, and for the rest of the time the muons are surrounded by spin-singlet dimers, so they do not feel $\delta$ and no depolarization occurs.  The fraction $f$ can also be interpreted as the relative hopping frequency of unpaired spins from one magnetic site to another \cite{Uemura_SCGO,Bono,Bono_2005}.  This model is in remarkable coincidence with the situation predicted by the theory for the oxynitride \cite{Uematsu_PRL}, where orphan spins which failed to find dimer pairs or unpaired spins generated by unconfined spinon excitation are expected to show up sporadically near the muon site for a mean fraction time $ft$, exerting the hyperfine field $\delta$ (with a fluctuation rate $\nu$) to the muon. The model is effectively equivalent with the case where all the frequency parameters in the uGKT function are scaled by the factor $f$ instead of $t$ \cite{Uemura_SCGO},    
   \begin{equation}
   	\begin{aligned} 
   		G_z^{\rm KT}(ft; H_L, \delta, \nu)  = G_z^{\rm KT}(t; fH_L, f\delta, f\nu).
   	\end{aligned}
   	\label{eqn:fDelta1}   	
   \end{equation} 
Here, the physical scenario is that the effective Mo hyperfine field and fluctuation rate are interpreted as $f$ times diluted by the time average. 
 
Meanwhile, depolarization due to spin fluctuations dominated by the Markovian process is described by the GKT function (corresponding to the limit of $f=1$ in the uGKT function).  In the case of fast fluctuation, we have
\begin{equation}
G_z^{\rm KT}(t; H_L, \delta, \nu^\prime)\simeq\exp(-\lambda^\prime t),\label{Gexp}
\end{equation}
\begin{equation}
\lambda^\prime \simeq\frac{2\delta^2\nu^\prime}{\gamma_\mu^2H_L^2+\nu^{\prime2}}\label{BPP}
\end{equation}
as an approximation of the GKT function ($\nu^\prime\gg\delta$). Note that when the fluctuation rate satisfies the condition $\nu^\prime\gg\gamma_\mu H_L$, Eq.~(\ref{BPP}) is least dependent on $H_L$. (The contributions of Mo$^{4+}$ domains from Mo$^{5+}$ domains to paramagnetic fluctuations are not distinguished, and $\delta$ is taken as the mean value.)  Given that the long-tail part of $A(t)$ has least dependence on $H_L$, we presume that Eq.~(\ref{BPP}) is a good approximation for this part.

The time-dependent asymmetry can be then analyzed by curve-fitting using the following function,  
  \begin{equation}
  	\begin{aligned} 
 		A(t) =& A_0[x\cdot G_z^{\rm KT}(t; fH_L, f\delta, f\nu)\\+&(1-x)\cdot\exp(-\lambda^\prime t)]+A_{\rm b}.
  	\end{aligned}
  	\label{eqn:asy_ft}   	
  \end{equation}  
where $x$ is the volume fraction where the muon senses short-time sporadic Mo hyperfine field fluctuations during excitation, and $1-x$ is that for the rest of the region where the muon senses spin fluctuations due to continuous thermal processes.  (For the undecouplable character of the time spectra in Lu$_2$Mo$_2$O$_{5-y}$N$_2$, see Appnd. B.)

Following previous examples of analysis using uGKT functions in CuAl$_2$O$_4$ \cite{Cho} and Cr-kagome bilayer compounds \cite{Bono}, we assumed that $x$ changes with LF as the effect of LF on the spin dynamics [$x=x(H_L)$ ].
Note that the suppression of spin fluctuations near the critical slowing down by applying external magnetic field is a well-known phenomenon.  In the least-square analysis, the parameters other than $x$ were assumed to be independent of LF, and fixed to values obtained from ZF-$\mu$SR data at each temperature point.  A curve-fitting analysis based on these assumptions yielded excellent least-squares fits for the entire set of ZF/LF-$\mu$SR time spectra.

\subsection{The local spin dynamics in each type of domains}

\begin{figure}[t]
  	\begin{center}
  		\includegraphics[width=0.48\textwidth]{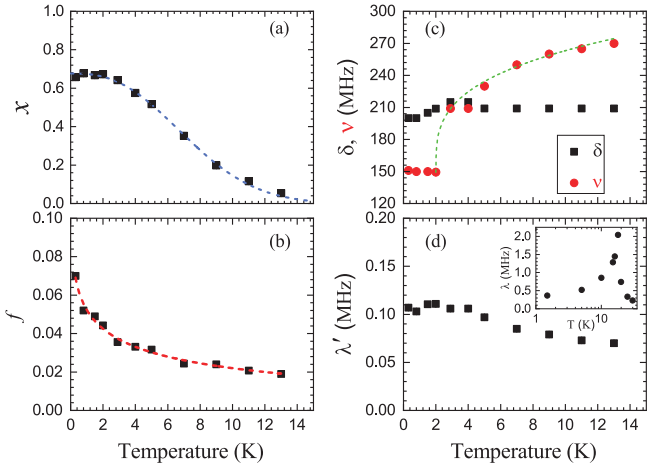}
  	\end{center}
  	\caption{The volume fraction $x$ at ZF (a), scaling factor $f$ (b),  hyperfine parameter $\delta$ and fluctuation rate $\nu$ (c) for the domains of sporadic fluctuation over the temperature range 0.3--13 K. For the dashed curves in (a)--(c), see text.  The relaxation rate $\lambda^{\prime}$ in the paramagnetic domains is shown in (d), where the inset shows the relaxation rate in Lu$_2$Mo$_2$O$_7$ quoted from Ref.~\cite{ClarkThesis}.}
  	\label{Heliox_ZF_parm}
  \end{figure}

 The temperature dependence of the parameters in Eq.~(\ref{eqn:asy_ft}) deduced from the curve fit is summarized in Fig.~\ref{Heliox_ZF_parm}. For the sporadic fluctuation component in Fig.~\ref{Heliox_ZF_parm}(a), a plateau is observed for $x$ below $\sim$2 K (with $x=x_0\simeq0.67$). Similarly, the hyperfine field fluctuation in Fig.~\ref{Heliox_ZF_parm}(c) also levels off at $\nu=\nu_0\sim$150 MHz after decreasing rapidly with decreasing temperature toward 2 K. The dashed curve is the result of fitting to the relation $\nu = \nu_0 + c\cdot(T-T_p)^\alpha$, with $\alpha = 0.30(3)$ for $T_p=2$ K. Such a behavior of $\nu$ is markedly different from the conventional magnetism where $\nu_0=0$; $\alpha$ is also unusually small (e.g., $\alpha=3$ for spin-wave excitation).  Since $\nu/\delta\simeq0.75$, the upward shift in the long-tail part of $A(t)$ with increasing LF seen in Fig.~\ref{LFtspec}(a)-(c) can be attributed to the component represented by the uGKT function. The hyperfine field ($\delta\simeq200$--210 MHz) is comparable to $\nu$ up to 13 K, which is the main reason why this part of the relaxation cannot be described by Eq.~(\ref{BPP}), which is valid only for the case $\nu\gg\delta$. 

At higher temperatures above $\sim$2 K, $x$ decreases monotonically and approaches zero around 13 K. This suggests that the magnetic domains exhibiting sporadic fluctuations are changing to those dominated by Markovian fluctuations. The sharp decrease of $f$ with increasing temperature for 0.3--2 K seen in Fig.~\ref{Heliox_ZF_parm}(b) also suggests that the sporadic fluctuations are strongly suppressed by thermal excitation. The anti-correlation between $f$ and $\nu$ at temperatures above $\sim$2 K is consistent with the interpretation that sporadic field fluctuations are a hallmark of the quantum fluctuations.  

Magnetic domains dominated by the Markovian fluctuations also exhibit qualitatively different behavior than normally be expected.  While $\lambda^{\prime}$ gradually increases down to $\sim$2 K  [see Fig.~\ref{Heliox_ZF_parm}(d)], there is no cusp-like peak signifying spin glass freezing in the system is observed (see the cusp-like peak in the muon depolarization rate near $T_g\simeq 16$ K seen in Lu$_2$Mo$_2$O$_7$ in the inset \cite{ClarkThesis}). The fact that this does not occur is also evident in the ZF-$\mu$SR spectra in Fig.~\ref{ZF_spectra_Heliox}, where there is  no drastic change near $T_g$. In addition, the least dependence of $\lambda^{\prime}$ on $H_L$ (see below) indicates that it is in the relatively fast fluctuation limit down to 0.3 K. Assuming $\delta\simeq1.73$--$2.83\delta_{\rm e}$ (for $S=1/2$--1),  the fluctuation rate can be estimated as $\nu^{\prime}\simeq\delta^2/\lambda^{\prime}\simeq10^{6}$--$10^{7}$ MHz,
 which is orders of magnitude faster than $\nu$ for the sporadic fluctuation. Thus, the transition to the quasi-static spin glass is suppressed even in the paramagnetic domains. 
 
\begin{figure}[t]
   	\begin{center}
   		\includegraphics[width=0.48\textwidth]{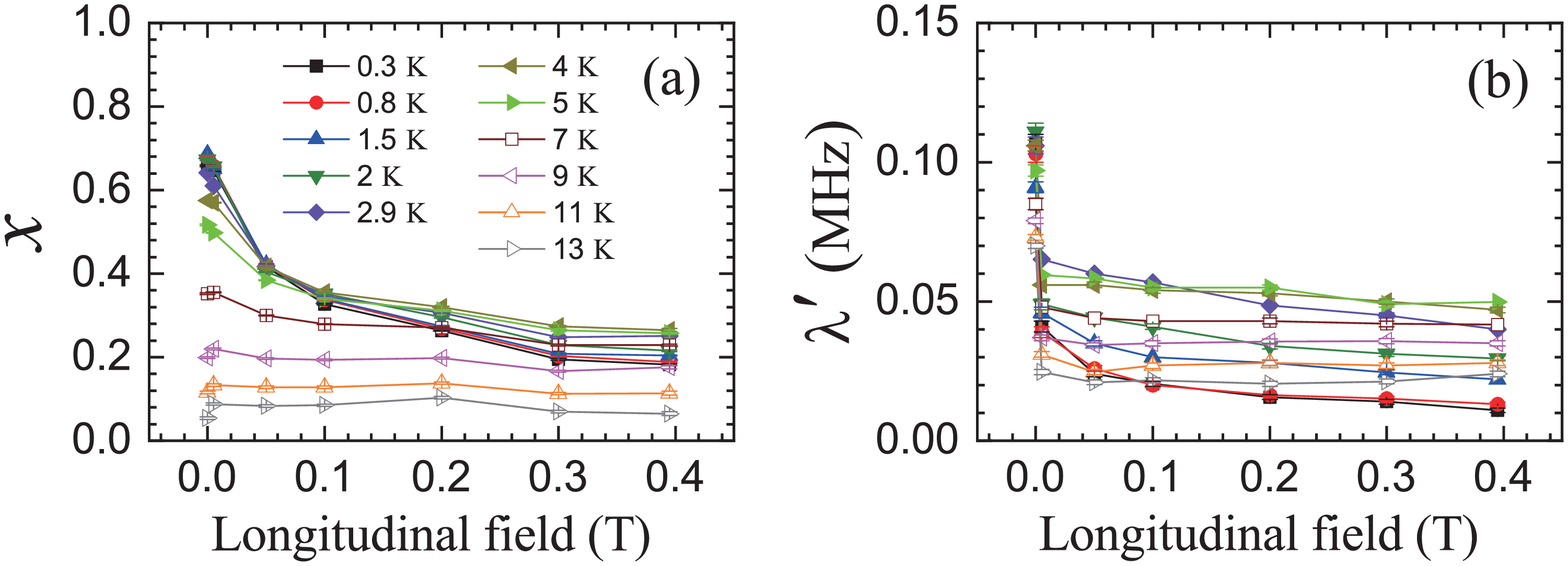}
   	\end{center}
   	\caption{
   			Longitudinal field (LF) dependence of (a) $x$, and (b) $\lambda^{\prime}$ deduced from curve-fits of the LF-$\mu$SR time spectra. Solid lines are guide to the eye. }
   	\label{x_lambda_LF}
   \end{figure}

Now, let us briefly examine the effect of the external magnetic field on spin dynamics, which is summarized in Fig.~\ref{x_lambda_LF}. The volume fraction $x$ at low temperatures decreases significantly with increasing field around $H_L\sim$50 mT and approaches a constant value $x/x_0\sim0.5$ towards 0.4 T [see Fig.~\ref{x_lambda_LF}(a)], which is common to CuAl$_2$O$_4$ \cite{Cho} and SCGO(p) \cite{Bono}. In contrast, $\lambda^\prime$ for the Markovian spin fluctuation is mostly independent of $H_L$; the sharp decrease seen from 0 to 5 mT is due to the decoupling of the nuclear dipolar fields ($\delta_{\rm n}$). This is in line with the estimation that $\nu^\prime\gg\gamma_\mu H_L$ ($\le3.4\times10^2$ MHz  for the present experiment). 

\section{Discussion}

Since the present oxynitride sample consists of a mixture of Mo$^{4+}$ and Mo$^{5+}$ ions, the observation of two magnetic domains with different dynamic properties is presumably due to the heterogeneous distribution of Mo valence;  the paramagnetic fraction as large as $1-x\gtrsim0.33$ seems too large to be attributed to the spin-triplet component induced by the bond randomness \cite{Uematsu_PRL}. Therefore, it is reasonable to attribute the sporadic fluctuation component to the Mo$^{5+}$-rich domains and the continuous fluctuation component to the Mo$^{4+}$-rich domains.  The saturation of $x$ below $\sim$2 K at about $x_0\simeq0.67$ can be interpreted as the net domain size for the sporadic fluctuations being limited by the fraction of Mo$^{5+}$ ions.  From this, it is inferred that $x_0$ is determined by oxygen deficiency, i.e., $x_0\simeq1-y$.

The above conclusion is also supported by the magnitude of the hyperfine parameter for the sporadic fluctuation. Using the calculated value $\delta_{\rm e}/\mu_B$ in Fig.~\ref{Musite}(c), we obtain the effective moment size $\mu_{\rm eff}=\delta/\delta_{\rm e}=1.63$--$1.71\mu_B$. This value is close to $\mu_{\rm eff}=1.73\mu_B$ for $S=1/2$ in agreement with that for the Mo$^{5+}$ ions. Considering that $\delta_{\rm e}$ of the paramagnetic state is dominated by the sum of the hyperfine field $\delta_{\rm Mo}$ exerted by the four most adjacent Mo spins ($\delta_{\rm Mo}\simeq\frac{1}{2}\delta_{\rm e}=61.5$ MHz/$\mu_B$, from the relation $\delta_{\rm e}^2\simeq4\delta_{\rm Mo}^2$), the  active spin number is estimated to be $\delta/\delta_{\rm Mo}=3.3$--3.4 per muon site. This can be interpreted as corresponding to a local cluster of spins generated by the sporadic breakdown of spin-singlet correlations; $\delta$ could have been deduced to $\sim$$\frac{1}{2}\delta_{\rm e}$ for the isolated spinon excitations. (Note that this does not necessarily imply the absence of spinon excitations, but only that time scale of the spinon fluctuations is outside the $\mu$SR time window.)

However, the following two facts must be considered.
1) $x$ shows a strong temperature dependence, which is not expected if $x$ is simply determined by the ratio of Mo$^{5+}$ to Mo$^{4+}$.
2) Spin glass freezing, which is expected to occur in the Mo$^{4+}$ domain, is not observed in the present sample. Therefore, we discuss the possibility that these are due to the effect of bond randomness.

First, given that the sporadic fluctuations are associated with the QSL state, $x$ can be regarded as the fractional volume of the ``QSL phase''.  Moreover, the gradual decrease of $\nu$ with decreasing temperature and leveling-off below $\sim$2 K suggests a critical behavior associated with some kind of the second-order phase transition to the QSL state. Therefore, we postulate that $x$ is an order parameter of the QSL state similar to the superfluid density in the superconducting phase transition.  
The temperature dependence of $x$ then suggests the energy gap for breaking the QSL state exists at an energy scale of $\Delta_0\sim10^1$ K ($\sim$1 meV).

The complication here is that $\Delta_0$ can be inhomogeneous due to bond randomness modulating the exchange interaction $J_{ij}$ between singlet pairs $S_i$-$S_j$; the Mo-O-Mo paths is disrupted by N substitution for O \cite{Clark_JSSC}. According to the random-bond spin-1/2 isotropic Heisenberg model on a pyrochlore lattice with antiferromagnetic nn interactions, the introduction of a bond-independent uniform distribution of the normalized $J_{ij}$ between $[1-\varepsilon,1+\varepsilon]$ (with $0\le\varepsilon\le1$) leads to the disappearance of the excitation gap and the emergence of the spin-triplet states for $\varepsilon>\varepsilon_c\simeq0.6$ \cite{Uematsu_PRL}. This result suggests that $\Delta_0$ has a spatially non-uniform distribution to accompany nodal (null-gap) parts, which is somewhat similar to the  superconducting gap with nodes. 

In order to examine the effect of bond-randomness on the observed gap more quantitatively, let us consider the possibility of spin-singlet correlations in the spinon Fermi surface as the origin of this gap, and further assume that $\Delta_0$ is the BCS gap with the transition temperature $T_c$ ($=\Delta_0/1.764k_B$).  Since the variation of $\Delta_0$ with temperature ($T$) can be approximated by $[1-(T/T_c)^4]$ \cite{Tinkham:04}, the $T$ dependence of $x$, taking into account the effect of randomness on the Gaussian distribution, would be given by,
\begin{equation}
x \propto\int_0^\infty T_c\left[1-\left(\frac{T}{T_c}\right)^4\right]P(T_c)dT_c, 
\end{equation}
\begin{equation}
 P(T_c) = \frac{1}{\sqrt{2\pi}T_\varepsilon}e^{-\frac{(T_c-T_0)^2}{2T_\varepsilon^2}},\nonumber
\end{equation}
where $T_0$ and $T_\varepsilon$ are the mean and the standard deviation of the distribution for $T_c$. Curve fitting $x$ in this form yields $T_0=7.20(7)$ K [i.e., the mean value of $\Delta_0$ is 1.10(1) meV] and $T_\varepsilon = 4.03(5)$ K [$T_\varepsilon/T_0=0.560(5)$], where the dashed curve in Fig.~\ref{Heliox_ZF_parm}(a) shows the result. In other words, the states of about 3.6\% [$=\int_{-\infty}^0 P(T_c)dT_c$] correspond to those without gap. 

This scenario is supported by the fact that the magnitude of $f$ is close to the ratio $R$ of the samples with spin-triplet states inferred from the theoretical simulation for the oxynitride \cite{Uematsu_PRL}. $R$ becomes comparable to $f$ when the relative magnitude of bond-randomness exceeds a threshold ($\epsilon\ge\epsilon_c$) where the excitation gap vanishes.  From these observations, it is reasonable to suppose that the sporadic fluctuations observed in $\mu$SR correspond to gapless excitation associated with such gap nodes.  This is in line with the inference of gapless excitations from INS experiment \cite{Clark_PRL}, while the wide distribution of $\Delta_0$ would make it difficult to identify features associated with $\Delta_0$ from INS spectra.

Experimentally, it has been reported that $f$ is proportional to the fractional concentration $p$ of Cr spins ($f\propto p$) in kagome bilayer compounds, SrCr$_{9p}$Ga$_{12-9p}$O$_{19}$ and Ba$_2$Sn$_2$ZnGa$_{10-7p}$Cr$_{7p}$O$_{22}$ \cite{Bono}.  This indicates that $f$ is reduced by the random substitution of Cr with Ga (i.e., site-randomness), suggesting that $f$ can be also regarded as a measure for the coherence of the QSL state.  The decrease in $f$ with increasing temperature in Fig.~\ref{Heliox_ZF_parm}(b) can be interpreted as the coherence also being destroyed by thermal excitation. A curve fit in the form $f=f_0\exp[-(T/T_f)^\beta]$ yields $f_0= 0.14(5)$ and $T_f = 0.97(1.17)$ K with $\beta = 0.27(7)$. The relatively small value of $k_BT_f$ compared with $\Delta_0$ and the small $\beta$ ($\ll1$) indicate that $\Delta_0$ has a broad distribution. This may correspond to the case where the distribution width $\varepsilon$ is greater than $\varepsilon_c$  in the aforementioned theory.


Finally, we would like to comment on the the fact that the magnetic domain dominated by fast paramagnetic fluctuations (mainly composed of Mo$^{4+}$ ions) do not undergo a spin freezing. Considering that Lu$_2$Mo$_2$O$_7$ ($S=1$) exhibits spin glass freezing at $T_g \simeq16$ K \cite{Clark_JSSC}, it suggests that the glass transition occurring in pure Lu$_2$Mo$_2$O$_7$ may also be suppressed by bond randomness. This is seemingly at odds with the consensus that the origin of the spin glass transition lies in some randomness (e.g., a quenched disorder in atomic positions that has been favored for Y$_2$Mo$_2$O$_7$), and it may provide a clue to the mystery of the spin glass transition in Mo oxide pyrochlore.

\subsection{Acknowledgements}
We thank helpful discussions with H. Kawamura and M. J. P. Gingras.  Thanks are also to the MLF staff for their technical support for the muon experiment, which was conducted under the support of Inter-University-Research Programs (IURP, Proposals No.~2020B0388, 2020MI21, and 2021MI21) by Institute of Materials Structure Science (IMSS), KEK.  XANES experiments were conducted under the support of IURP (Proposals No.~2021V004) by IMSS, KEK.  The neutron scattering experiment received approval from the Neutron Scattering Program Advisory Committee of IMSS, KEK (Proposal No. 2019S06). S.K.D. also acknowledges support from Y.~Nagatani, and K.~Shimomura.  Thanks are also to H.~Lee for helping DFT calculations. This work was partly supported by the JSPS KAKENHI Grant No.~16H06439 and by the Elements Strategy Initiative to Form Core Research Centers, from the Ministry of Education, Culture, Sports, Science, and Technology of Japan (MEXT) under Grant No.~JPMXP0112101001.

\vspace{0.5in}

\noindent
\section{APPENDICES}

\setcounter{figure}{0}
\setcounter{table}{0}
\setcounter{equation}{0}
\renewcommand{\thefigure}{A\arabic{figure}}
\renewcommand{\thetable}{A\arabic{table}}
\renewcommand{\theequation}{A\arabic{equation}}

\subsection{Appendix A. Sample Characterization}

\begin{figure}[h]
	\begin{center}
		\includegraphics[width=0.42\textwidth]{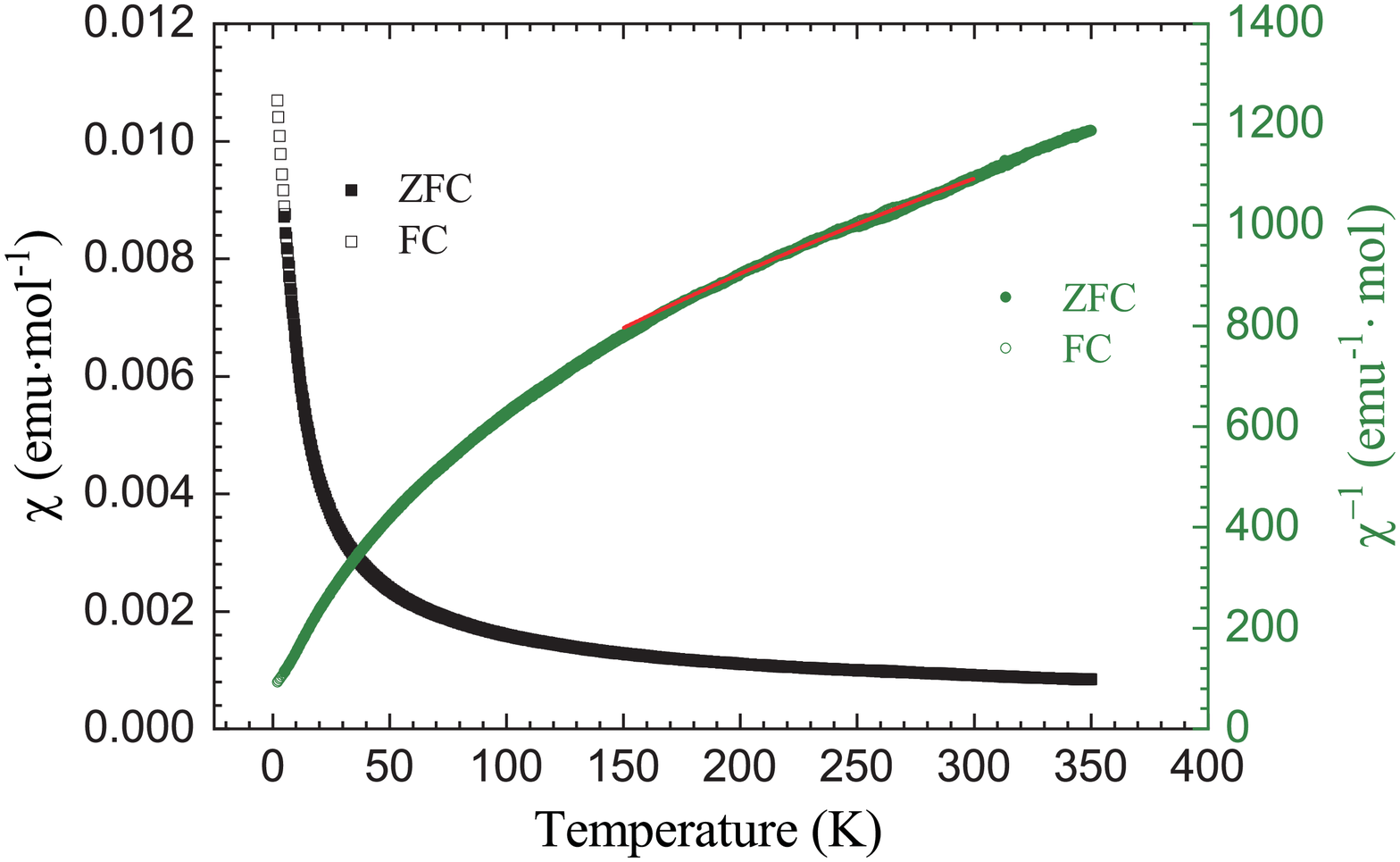}
	\end{center}
	\caption{The magnetic and inverse susceptibility ($\chi$) of
Lu$_2$Mo$_2$O$_{5-y}$N$_2$ with a Curie-Weiss fit to the high temperature region of $\chi^{-1}$ shown by the solid line. (See Table \ref{CWfits} for the numerical result.)}
	\label{Chi}
\end{figure}

Magnetic susceptibility ($\chi$) of Lu$_2$Mo$_2$O$_{5-y}$N$_2$ was measured using a SQUID magnetometer (Quantum Design Co.) at an applied field of 0.1 T from 2 K to 350 K in both zero-field-cooled (ZFC) and field-cooled (FC) conditions.  As shown in Fig.~\ref{Chi}, there is no significant difference between ZFC and FC data, which is consistent with the absence of  spin freezing. The effective magnetic moment $\mu_{\rm eff}$ and Weiss constant extracted from the Curie-Weiss curve fitting depend on the temperature range over which the fitting is performed.  The results are summarized in Table \ref{CWfits}. The values of $\mu_{\rm eff}$ are intermediate between those for Lu$_2$Mo$_2$O$_5$N$_2$ ($\sim$1.1$\mu_B$) and Lu$_2$Mo$_2$O$_7$  ($\sim$1.9$\mu_B$) reported in Ref.~\cite{Clark_PRL}, and the average valence was estimated to be Mo$^{4.6(2)+}$.  

\begin{table}[b]
\begin{center}
\begin{tabular}{c|cc|cc}
\hline\hline
\multirow{2}{*}{Fit region (K)} & \multicolumn{2}{c|}{This work} & \multicolumn{2}{c}{L. Clark {\it et al.} \cite{Clark_PRL}} \\
 & $-\theta_{\rm CW}$ (K) & $\mu_{\rm eff}$  ($\mu_B$) &  $-\theta_{\rm CW}$ (K) & $\mu_{\rm eff}$  ($\mu_B$) \\
 \hline
 150--300 & 115(10) & 1.41(6) &121(1)  &1.11(1)\\
 200--300 & 135(22) & 1.47(8) & 135(1) &1.13(1) \\
 250--300 & 148(2) &1.536(5) & 152(2) &1.16(1)\\
 \hline \hline
\end{tabular}
\caption{Result of the Curie-Weiss fit to Lu$_2$Mo$_2$O$_{5-y}$N$_2$. The corresponding result for the Lu$_2$Mo$_2$O$_5$N$_2$ sample reported in Ref.\cite{Clark_PRL} are quoted for comparison.
}\label{CWfits}
\end{center}
\end{table}

In the CHN elemental analysis, a sample on an alumina sample boat placed in a quartz tube was combusted at $\sim$1000 $^\circ$C, decomposed in a mixture of He (150 ml/min) and O$_2$ (20 ml/min) gases and allowed to pass through a copper oxide layer filled as an oxidant to detect H as H$_2$O, C as CO$_2$, and N as the remaining gas component to be detected. The analysis showed that the N content was 4.74(30) wt\%. This is in good agreement with the theoretical value of 4.31 wt\%, and it is concluded that N is contained in almost the same composition as the chemical formula. 

Powder XRD data were collected at room temperature using the BL02B2 diffractmeter at SPring-8 with a wavelength of 0.0420526(1) nm.  The sample ($\sim$10 mg) was sealed in a glass capillary with an inner diameter of 0.2 mm under atmospheric pressure conditions, where a volumetric filling factor was approximately 50\%. The exposure time was 20 seconds, and the capillary was measured while rotating.  Rietveld refinement on the cubic $Fd\bar{3}m$ model was performed using the JANA2006 code package \cite{Petricek:14}, and analysis results in perfect agreement with the pyrochlore structure were obtained.

\begin{figure}[t]
	\begin{center}
	\includegraphics[width=0.4\textwidth]{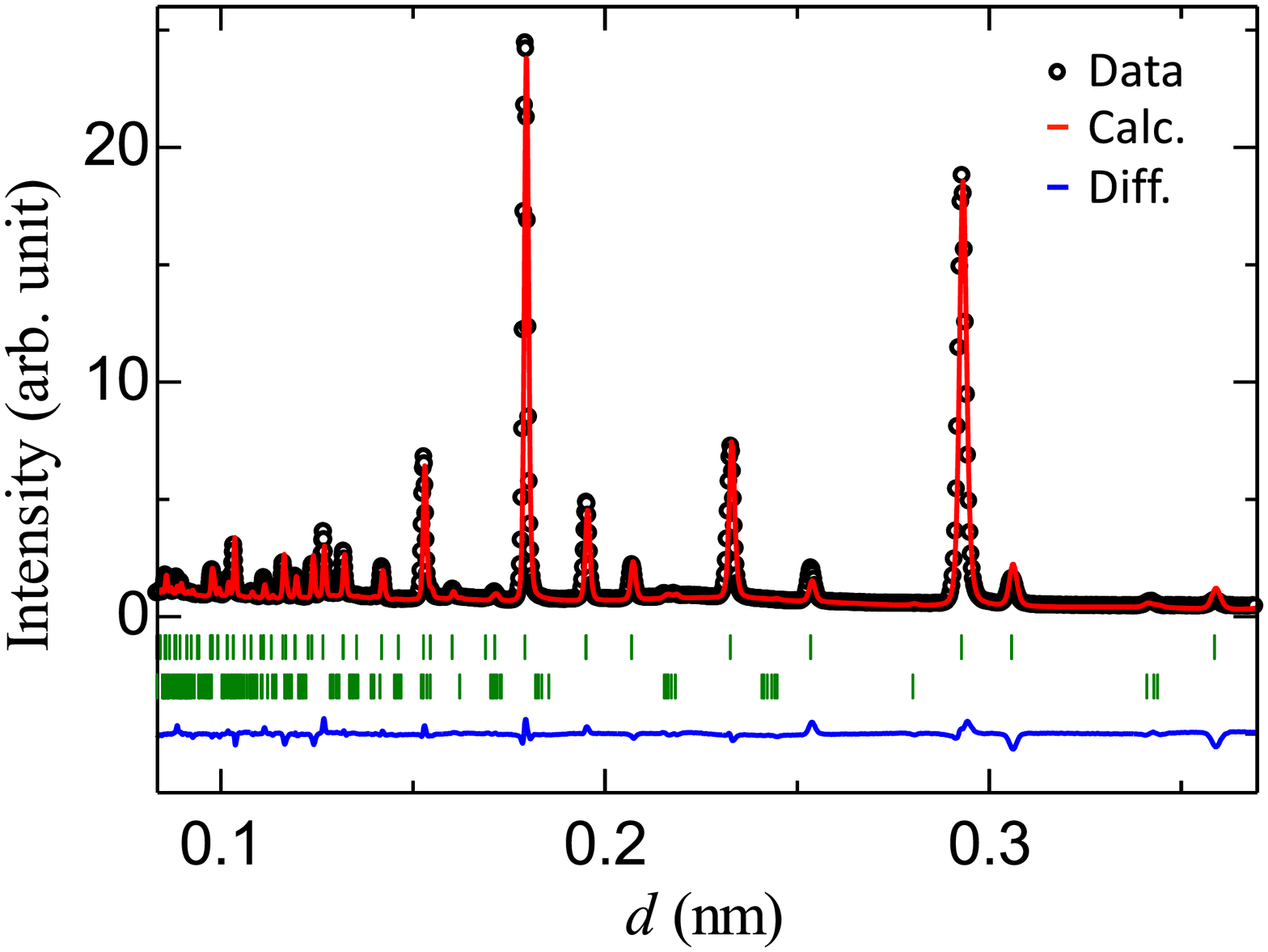}
	\end{center}
	\caption{Powder diffraction profiles of oxynitride sample with the cubic $Fd\bar{3}m$ model. Rietveld refinement results: observed (circles), calculated (line), and residual (line below the vertical bars) diffraction profiles. Top tick marks show the reflection positions for the oxynitride pyrochlore phase and bottom tick marks a MoO$_2$ impurity phase ($\sim$1\%) that was present in the oxide precursor.}
	\label{pND}
\end{figure}

\begin{table}[b]
\begin{center}
\begin{tabular}{c|cccccc}
\hline\hline
Atom & Site & $x$ & $y$ & $z$ & Occupancy & $U_{\rm iso}$ \\
 \hline
 Lu &16$d$ & 1/2 & 1/2 & 1/2 & 1.0 & 0.02822(8)\\
 Mo & 16$c$  & 0 & 0 & 0 & 1.0 & 0.03877(9)\\
O/N & 48$f$ &  0.33709(1) & 1/8 & 1/8 & 0.5830(3)/0.3211(2) & 0.04547(6)\\
O'/N' & 8$b$ & 3/8 &  3/8 &  3/8 & 0.9574(11)/0.0 & 0.01729(15)\\
 \hline \hline
\end{tabular}
\caption{Refined atomic coordinates and occupancies for Lu$_2$Mo$_2$O$_{5-y}$N$_2$ [$a = 1.0165750(8)$ nm]. Total $R_{\rm wp} = 6.70$\%, $R_{\rm p} = 5.00$\%.}
\label{pNDTbl}
\end{center}
\end{table}

More detailed structural analysis, including that for oxygen deficiency, was performed by PND experiments using the BL21 instrument NOVA in the Material and Life Science Experimental Facility (MLF) at J-PARC. The sample ($\sim$0.45 g) was placed in a vanadium container with an outer diameter of 3.0 mm and a thickness of 0.1 mm. Diffraction data were collected at room temperature and a data collection time was 4 hours. The consistency of scattering data was checked via reference materials such as silicon powder (NIST SRM 640d). The cubic model was refined by the Z-Rietveld code \cite{Oishi:09,Oishi:12} for data collected over a lattice spacing of 0.083--0.373 nm on the 90$^\circ$ detector bank. A plot of the fit to the 90$^\circ$ data set is shown in Fig.~\ref{pND}. The refinement results are in good agreement with the pyrochlore structure, yielding the oxygen/nitrogen occupancies at the 48$f$ and 8$b$ sites as summarized in Table~\ref{pNDTbl}.  From these values, the oxygen deficit is estimated to be $y=0.455(3)$ and the Mo valence to be Mo$^{4.545(3)+}$. However, the O-site disorder characteristic of pyrochlore oxides was found to add considerable ambiguity into the analytical results, and the oxygen deficit was finally evaluated to be $y=0.5(1)$, corresponding to Mo$^{4.5(1)+}$.

The XANES spectra were measured at NW-10A of the Advanced Ring in the Photon Factory (PF-AR) at KEK, Japan to estimate the Mo valence in the oxynitride sample by comparison with those for Mo, MoO$_2$ and MoO$_3$ which contain Mo with valence of 0, $4+$ and $6+$, respectively. Normalization of the raw spectra was performed by subtracting linear pre-edge and polynomial post-edge backgrounds by Athena program \cite{Athena_program}. 

\begin{figure}[t]
	\begin{center}
		\includegraphics[width=0.38\textwidth]{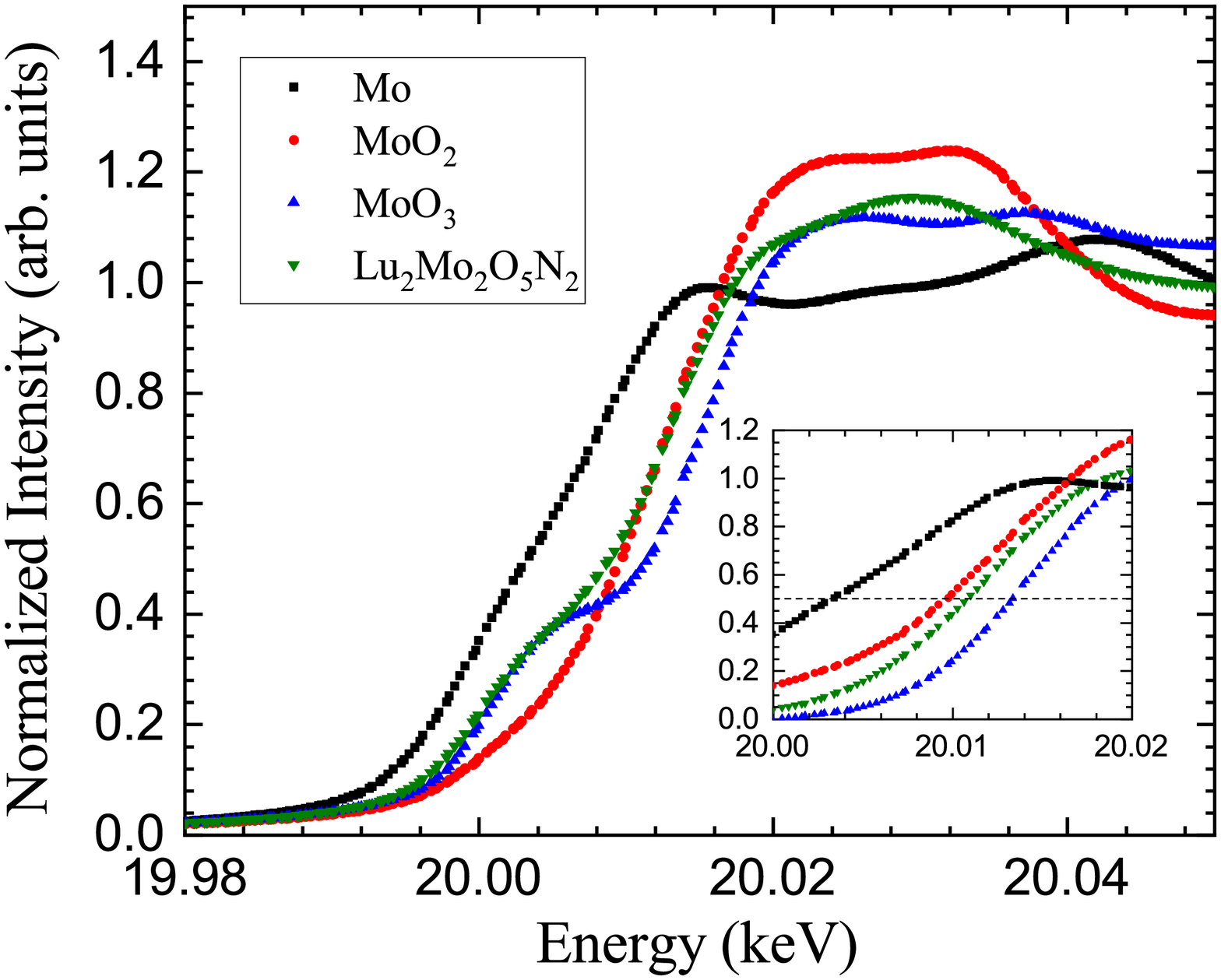}
	\end{center}
	\caption{XANES spectra around the Mo $K$-edge obtained for the oxynitride sample and reference compounds (Mo, MoO$_2$, and MoO$_3$). Inset: the spectra around the main edge after subtracting pre-edge peak near 20.003 keV.}
	\label{XANES}
\end{figure}

Normalized XANES spectra of Mo $K$-edge are shown in Fig.\ref{XANES}. The spectra of MoO$_3$ and oxynitiride samples exhibit a pre-edge peak around 20.003 keV, which is attributed to the transition from $1s$ to $4d$-$5p$ due to the $d$-$p$ orbital mixing. The absence of this feature for MoO$_2$ may be due to differences in local structural distortions; both MoO$_3$ and oxynitiride have considerable distortions in the MoO$_6$ octahedra, which may lead to stronger $d$-$p$ mixing. To compare the main edges, the spectra are shown with the pre-edge contribution subtracted using a curve-fit by  Gaussian peaks (see the inset of Fig.~\ref{XANES}).  The energy of the Mo $K$-edge, defined by the middle of the normalized intensity in these corrected spectra, suggests that the average valence is Mo$^{4.6(1)+}$.

\setcounter{figure}{0}
\setcounter{table}{0}
\setcounter{equation}{0}
\renewcommand{\thefigure}{B\arabic{figure}}
\renewcommand{\thetable}{B\arabic{table}}
\renewcommand{\theequation}{B\arabic{equation}}

\subsection{Appendix B. $\mu$SR Experiment and DFT calculation}


In the usual $\mu$SR experiment, the time evolution of muon spin polarization is observed by monitoring the time-dependent asymmetry of positrons emitted from muons upon beta-decay (the mean lifetime $\tau_{\mu}\sim$ 2.198 $\mu$s). The time-dependent positron counting rate is given by \cite{Rotaru_PRB}
\begin{equation}
  N(t) = N_0 e^{-t/\tau_{\mu}}[1+A(t)]+B
\end{equation}  
where, $N_0$ is the initial rate, $A(t)$ is the muon-positron decay asymmetry, and $B$ is the time-independent background. More specifically, the muon polarization along the initial muon spin direction [$G_z(t)$, where $z$ is parallel to the beam direction] is observed by a set of positron counters placed in the forward and backward positions [$N_{\rm B}(t)$ and $N_{\rm F}(t)$] against sample \cite{Undecouplable_Gaussian_Lago_2005},
\begin{equation}
	\begin{aligned}
	A(t) = \frac{N_{\rm B}(t)-\alpha N_{\rm F}(t)}{N_{\rm B}(t)+\alpha N_{\rm F}(t)}= A_0 G_z(t) +A_{\rm b}	
\label{eqn:A_FB_alpha}
	\end{aligned}
\end{equation}
where $\alpha$ is the factor to correct the instrumental asymmetry due to the difference in the efficiencies between the forward and backward counters, determined from the spectra obtained under a weak transverse field in samples in non-magnetic phase. $A_0$ is the initial muon asymmetry for the sample and $A_{\rm b}$ is the background from the muons which missed the sample. The magnitude of $A_{\rm b}$ depends on the sample environment including the cryostat;  $A_{\rm b}$ comes mainly from the sample holder (made of silver) in the $^3$He cryostat.

\begin{figure}[t]
	\begin{center}
		\includegraphics[width=0.4\textwidth]{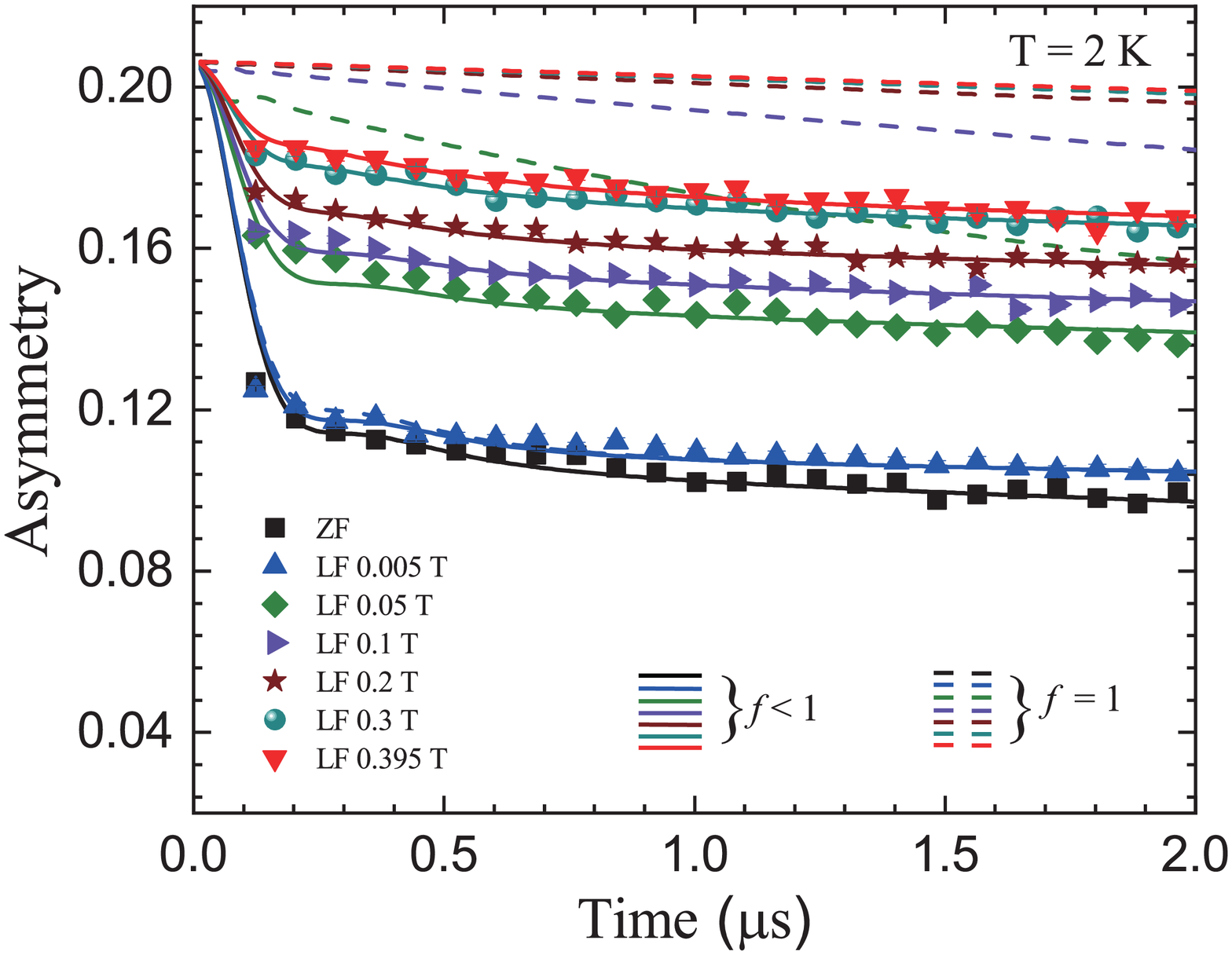}
	\end{center}
	\caption{$\mu$SR time spectra observed at 2 K in Lu$_2$Mo$_2$O$_{5-y}$N$_2$ under various longitudinal fields (LFs), where the solid curves represent fits with $f$ as a free parameter common to all LFs  (yielding $f=0.044$), and dashed curves represent $f=1$. The data points for $t<0.1$ $\mu$s were not used for analysis to mitigate the ambiguity due to the time resolution ($\simeq100$ ns). }
	\label{LFexpnd}
\end{figure}

In the first-round analysis of the $\mu$SR time spectra, we attempted a curve fit without the sporadic fluctuation model (i.e., corresponding to an analysis using Eq.~(\ref{eqn:asy_ft}) with $f=1$), which failed to reproduce the data. The dashed lines in Fig.~\ref{LFexpnd} show the field dependence of $A(t)$ corresponding to the internal magnetic field ($\delta/\gamma_\mu\sim$0.011 T) that is evaluated from the magnitude of depolarization ($\delta\simeq9.23$ MHz) at $t=0$--0.2$\mu$s. The data points show a much weaker field dependence than these, which is in close similarity with the ``undecouplable'' behavior observed earlier in various frustrating antiferromagnets \cite{Cho,Khuntia_PRL,Undecouplable_Gaussian_FUKAYA, Undecouplable_Gaussian_Aczel_PRB,Undecouplable_Gaussian_PRB_Yoon,Uemura_SCGO,Bono,Undecouplable_Gaussian_Lee_PRB}. From this result we conclude that a sporadic model is needed for the analysis of these spectra.

For the DFT calculations to estimate muon site, we employed the VASP code package \cite{VASP} to perform a total energy calculation for a unit cell (88 atoms) containing an interstitial hydrogen atom (to mimic muon) without structural relaxation, using the GGA-PBE exchange correlation function with cutoff energy 400~eV and 14$K$ points. We performed total energy calculations without N substitution (corresponding to  Lu$_2$Mo$_2$O$_7$) and for 16 of 56 O atoms randomly substituted with N and compared them, finding little difference near the energy minima.

\setcounter{figure}{0}
\setcounter{table}{0}
\setcounter{equation}{0}
\renewcommand{\thefigure}{C\arabic{figure}}
\renewcommand{\thetable}{C\arabic{table}}
\renewcommand{\theequation}{C\arabic{equation}}

\subsection{Appendix C. Hyperfine Interactions}
In the following, we summarize the hyperfine interactions between muon and nuclear/electron spins in Lu$_2$Mo$_2$O$_{5-y}$N$_2$ that is represented by the effective local magnetic field  ${\bm H}({\bm r})$ and the corresponding time variation of the muon spin polarization (time spectrum) to be observed.

Let us first consider the interaction between muon and nuclear spins.  In general, the term ``hyperfine interaction" includes both magnetic dipole interaction and the Fermi contact interaction. However, since both nuclei and muons are well localized in the ground state, the interaction between them is mainly magnetic dipole interaction. The Hamiltonian is then given as
\begin{eqnarray}
{\cal H}/\hbar &= & {\cal H}_{\rm Z}/\hbar+{\cal H}_{\rm n}/\hbar, \\
{\cal H}_{\rm Z}/\hbar & =& \gamma_\mu {\bm S}_\mu \cdot {\bm H}_0+\sum_{n,i} f_n\gamma_n {\bm I}_{ni}\cdot{\bm H}_0,\\
{\cal H}_{\rm n}/\hbar &=&\gamma_\mu{\bm S}_\mu\sum_{n,i} f_n\gamma_n\hat{A}_{\rm d}^{ni}{\bm I}_{ni},\label{nucldip}
\end{eqnarray}
where ${\cal H}_{\rm Z}$ represents the Zeeman interaction for muon and nuclear spins under an external magnetic field ${\bm H}_0$, ${\bm S}_\mu$ is the muon spin, $\gamma_\mu=2\pi \times 135.53$ MHz/T is the gyromagnetic ratio of muon spin, ${\bm I}_{ni}$ is the $n$th kind of nuclear spin at distance $r_{ni}$ ($n=1,2,3,$ and 4 for $^{175}$Lu, $^{95}$Mo, $^{97}$Mo, and $^{14}$N) on the $i$th lattice point, $\gamma_n$ and $f_n$ are the gyromagnetic ratio and natural abundance (or occupation) of the $n$th nuclear spin, $\hat{A}_{\rm d}^{ni}$ is the magnetic dipole tensor
\begin{equation} 
(\hat{A}_{\rm d}^{ni})^{\alpha\beta}
=\frac{1}{r^3_{ni}}(\frac{3\alpha_{ni} \beta_{ni}}{r^2_{ni}}-\delta_{\alpha\beta}) \:\:(\alpha,\beta=x,y,z),\label{diptensor}
\end{equation} 
representing the hyperfine interaction between muon and nuclear magnetic moments at  ${\bm r}_{ni} = (x_{ni},y_{ni},z_{ni})$ being the position vector of the nucleus seen from muon. In the case of zero-external field (${\cal H}_{\rm Z}=0$), the effective magnetic field expressed as
\begin{equation}
{\bm H}({\bm r})={\bm H}_{\rm n}=\sum_n f_n\gamma_n\sum_i\hat{A}_{\rm d}^{ni}\overline{{\bm I}}_{ni}\label{bdip_n}
\end{equation}
 is used to obtain the effective Hamiltonian
\begin{equation}
{\cal H}/\hbar =\gamma_\mu{\bm S}_\mu\cdot{\bm H}_{\rm n},
\end{equation}
and the time evolution of the muon spin polarization ${\bm P}(t) =\langle {\bm S}_\mu(0)\cdot{\bm S}_\mu(t)\rangle/|{\bm S}_\mu(0)|^2$ can be obtained analytically using the density matrix of the muon-nucleus spin system for a small number of nuclei  (where $\gamma_n\overline{{\bm I}}_{ni}$ is the effective magnetic moment considering the electric quadrupole interaction for ${\bm I}_{ni}\ge1$).

\begin{figure}[t]
	\begin{center}
		\includegraphics[width=0.45\textwidth]{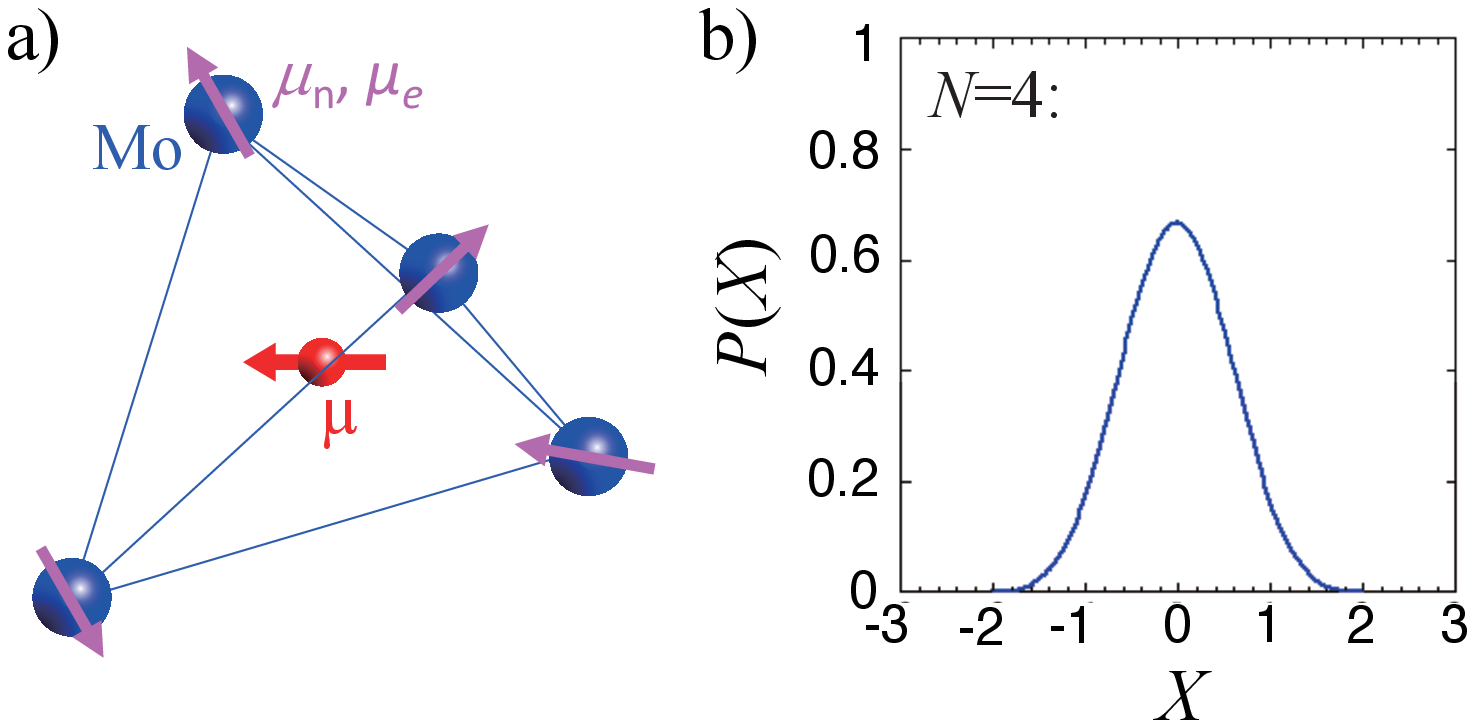}
	\end{center}
	\caption{(a) Schematic of a muon localized at the center of a Mo tetrahedron. The internal magnetic field at the muon site is governed by the magnetic dipole moment of the nuclear ($\mu_{\rm n}$) and electron ($\mu_{\rm e}$) moments. (b) Probability distribution $P(X)$ of random variable $X=\sum_i^N X_i$ generated by uniform random numbers $X_i=[-0.5,0.5]$ for $N=4$. }
	\label{hfi}
\end{figure}

On the other hand, if the coordination of the nuclear magnetic moment viewed from the muon is isotropic and the number of coordination $N$ is sufficiently large ($N\gtrsim4$), the classical spin treatment is easier, and the density distribution $P({\bm H})$ of ${\bm H}({\bm r})$ is approximated by a Gaussian distribution with zero mean value \cite{Hayano_PRB},
\begin{eqnarray}
P(H_\alpha)&=&\langle\delta(H_\alpha-H_\alpha({\bm r}))\rangle_{\bm r}\nonumber
\\&=&\frac{\gamma_{\mu}}{\sqrt{2\pi}\delta_{\rm n}}
\exp\left(-\frac{\gamma_{\mu}^{2}H_\alpha^{2}}{2\delta_{\rm n}^{2}}\right)  \:\:   (\alpha=x,y,z).
\label{ph}
\end{eqnarray}
To check the validity of this approximation, let us simulate the magnetic field exerted on the muon by a single magnetic dipole with random orientation at the $i$th site by a uniform random number $X_i=[-0.5,0.5]$. In this case, the magnetic field distribution from $N$ dipoles is represented as a distribution $P(X)$ of random variable $X=\sum_i^N X_i$. If the muon is at the center of the Mo tetrahedron as in Fig.~\ref{hfi}(a), then $P(X)$ takes the distribution shown in Fig.~\ref{hfi}(b) for $N=4$ which is in good agreement with Gaussian distribution. Note that this does not depend on whether the magnetic dipole is nuclear or electron-derived. $P(X)$ rapidly approaches the true Gaussian distribution with increasing $N$ (this is one example of what is called the "central limit theorem" in statistics). 

The magnitude of $\delta_{\rm n}$ is given by the second moment of ${\bm H}_{\rm n}$ as
\begin{equation}
\frac{\delta_{\rm n}^2}{\gamma_\mu^2}=\sum_n f_n\sum_i\sum_{\alpha,\beta}[\gamma_n(\hat{A}_{\rm d}^i)^{\alpha\beta}\overline{{\bm I}}_{ni}]^2, \label{delta_n}
\end{equation}
with $\beta$ taking all $x,y,z$, and the $\alpha$ over the $x,y$ components that are effective for longitudinal relaxation when $\hat{z}$ is the longitudinal direction; the $z$ component does not contribute to the relaxation because it gives a magnetic field parallel to the muon spin.
In this case, the time-dependent muon spin polarization ${\bm G}(t)$ is given by the motion of one muon spin projected onto ${\bm H}$ with the angle between the magnetic field ${\bm H}$ and the $\hat{z}$ axis as $\theta$,
\begin{equation}
P_z(t)=\cos^2\theta+\sin^2\theta\cos(\gamma_\mu Ht)\label{sz}
\end{equation}
which is averaged by $P({\bm H})$ in Eq.~(\ref{ph}) to yield the Gaussian Kubo-Toyabe function
\begin{eqnarray}
G_z^{\rm KT}(t;\delta_{\rm n})&=&\iiint_{-\infty}^{\infty}P_{z}(t)\Pi_\alpha P(H_{\alpha})dH_\alpha\nonumber\\
&=&\frac{1}{3}+\frac{2}{3}(1-\delta_{\rm n}^{2}t^{2})e^{-\frac{1}{2}\delta_{\rm n}^{2}t^{2}}. \label{gkt}
\end{eqnarray}
When $\delta_{\rm n}$ is relatively small, the recovery of polarization to 1/3 (for $t>\sqrt{3}/\delta_{\rm n}$) is out of the $\mu$SR time range ($\lesssim$20 $\mu$s), and we only observe the initial Gaussian damping, $G_z^{\rm KT}(t)\simeq \exp(-\delta_{\rm n}^2t^2)$ (which is the case for Lu$_2$Mo$_2$O$_{5-y}$N$_2$ at 298 K). The lineshape $G_z^{\rm KT}(t;H_L,\delta_{\rm n})$ under a longitudinal field  $H_L$ ($\parallel z$) can be derived by replacing $H_z$ in Eq.~(\ref{ph}) with $H_z-H_L$.

In the case of fluctuating $\delta_{\rm n}$ over time (e.g., due to the self-diffusion of muon), $G_z^{\rm KT}(t;H_L,\delta_{\rm n})$ is subject to the modulation and adiabatically approaches exponential decay with increasing fluctuation rate $\nu_{\rm n}$, where the detailed lineshape of $G_z^{\rm KT}(t;H_L,\delta_{\rm n},\nu_{\rm n})$ as a function of $\nu_{\rm n}$ and an external longitudinal field $H_L$ are found elsewhere \cite{Hayano_PRB}.

The magnitude of $\delta_{\rm n}$ is sensitive to the size of the nearest-neighbor nuclear magnetic moment $\gamma_n\overline{{\bm I}}_{ni}$ and the distance $|{\bm r}_{ni}|$ from the muon, and the position occupied by the muon as pseudo-hydrogen can be estimated by comparing the experimentally obtained $\delta_{\rm n}$ with the calculated value at the candidate sites. We used the DipElec code \cite{Kojima:04} to calculate $\delta_{\rm n}$ for Lu$_2$Mo$_2$O$_{5-y}$N$_2$.

In the paramagnetic state of Lu$_2$Mo$_2$O$_{5-y}$N$_2$, the hyperfine interaction between muon and Mo electronic spins are presumed to be dominated by magnetic dipolar interaction. (Because of the small electric charge of muon and its interstitial position, hybridization between the 1$s$ orbit of muon as pseudo-hydrogen and the outer-shell electrons of surrounding atoms is small to yield negligible Fermi contact interaction at the muon site.) 
The Hamiltonian for the magnetic interaction between muon and unpaired electron is then given by
\begin{eqnarray}
{\cal H}/\hbar &=& [{\cal H}_{\rm Z'}+{\cal H}_{\rm e}]/\hbar,\label{Htot}\\
{\cal H}_{\rm Z'}/\hbar & =& \gamma_\mu {\bm S}_\mu \cdot {\bm H}_0+\gamma_{\rm e}\sum_i{\bm S}_i \cdot {\bm H}_0 \\
{\cal H}_{\rm e}/\hbar &=&\gamma_\mu\gamma_e{\bm S}_\mu\sum_i\hat{A}_{\rm d}^{i}{\bm S}_{i}\label{HMu}
\label{Hspin}
\end{eqnarray}
where ${\cal H}_{\rm Z'}$ represents the Zeeman interaction for muon and electron spins, ${\cal H}_{\rm e}$ is the Hamiltonian of the muon interaction with $i$th electron spin ${\bm S}_i$, $\gamma_e$ is the gyromagnetic ratio of the electron ($=2\pi\times28.02421$ GHz/T). 
In the case of zero-external field (${\cal H}_{\rm Z'}=0$), the effective magnetic field is expressed as
\begin{equation}
{\bm H}({\bm r})={\bm H}_{\rm e}=\gamma_e\sum_i\hat{A}_{\rm d}^{i}{\bm S}_{i}.\label{bdip_e}
\end{equation}
By comparing Eq.~(\ref{bdip_n}) and Eq.~(\ref{bdip_e}), it is clear that the effective muon-electron interaction can be described by replacing nuclear magnetic moments with that of electron. This also means that ${\bm G}(t)$ must be described by the Gaussian Kubo-Toyabe relaxation function with the linewidth (=hyperfine parameter) replaced by
\begin{equation}
\frac{\delta_{\rm e}^2}{\gamma_\mu^2}=\sum_i\sum_{\alpha,\beta}[\gamma_e(\hat{A}_{\rm d}^i)^{\alpha\beta}{\bm S}_{i}]^2. \label{delta_e}
\end{equation}

Since we are only interested in the hyperfine parameter per a Bohr magneton ($S=1/2$) which is the measure for the effective magnetic moment $\mu_{\rm eff}$, it is straightforward to calculate $\delta_{\rm e}/\mu_{\rm eff}$ using Eq.~(\ref{delta_e}) which is implemented in the DipElec code \cite{Kojima:04}.

\bibliography{Mo-pyro-v2}
\end{document}